\documentclass[twocolumn,preprintnumbers,amsmath,amssymb,nobibnotes,nofootinbib,superscriptaddress]
{revtex4}

\usepackage{graphicx}
\usepackage{dcolumn}
\usepackage{color}
\RequirePackage[colorlinks=true
,urlcolor=blue
,anchorcolor=blue
,citecolor=blue
,filecolor=blue
,linkcolor=blue
,menucolor=blue
,linktocpage=true
,pdfproducer=medialab
]{hyperref}

\usepackage{mathrsfs}
\usepackage{bbm}

\newcommand{\mean}[1]{\langle#1\rangle}

\newcommand{\ov}[1]{\overline{#1}}
\newcommand{\hc}{\text{h.c.}}
\newcommand{\unity}{\mathbbm{1}}

\newcommand{\LL}{\mathscr{L}}

\def\cA{{\cal A}}

\def\cM{{\cal M}}

\def\cX{{\cal X}}

\def\cY{{\cal Y}}
\def\cy{{\bf y}}
\def\tr{{\rm Tr}}

\def\be{\begin{equation}}
\def\ee{\end{equation}}
\def\beq{\begin{equation}}
\def\eeq{\end{equation}}
\def\bc{\begin{center}}
\def\ec{\end{center}}
\def\bea{\begin{eqnarray}}
\def\eea{\end{eqnarray}}

\def\nn{\nonumber}

\begin{document}

%
%

\preprint{FTUAM-13-15} 
\preprint{IFT-UAM/CSIC-13-071}
\preprint{CERN-PH-TH/2013-146}
\preprint{DFPD-2013/TH/10}

\title{Leptonic Dynamical Yukawa Couplings}

\author{ R. Alonso}
\email{rodrigo.alonso@uam.es}
\affiliation{Instituto de F\'{\i}sica Te\'orica UAM/CSIC and Departamento de F\'isica Te\'orica,\\
Universidad Aut\'onoma de Madrid, Cantoblanco, 28049 Madrid, Spain}
\affiliation{CERN, Department of Physics, Theory Division CH-1211 Geneva 23, Switzerland}

\author{M. B. Gavela}
\email{belen.gavela@uam.es}
\affiliation{Instituto de F\'{\i}sica Te\'orica UAM/CSIC and Departamento de F\'isica Te\'orica,\\
Universidad Aut\'onoma de Madrid, Cantoblanco, 28049 Madrid, Spain}
\affiliation{CERN, Department of Physics, Theory Division CH-1211 Geneva 23, Switzerland}

\author{D. Hern\'andez}
\email{dhernand@ictp.it}
\affiliation{The Abdus Salam International Center for Theoretical Physics,\\
Strada Costiera 11, I-34151 Trieste, Italy}

\author{L. Merlo}
\email{luca.merlo@uam.es}
\affiliation{Instituto de F\'{\i}sica Te\'orica UAM/CSIC and Departamento de F\'isica Te\'orica,\\
Universidad Aut\'onoma de Madrid, Cantoblanco, 28049 Madrid, Spain}
\affiliation{CERN, Department of Physics, Theory Division CH-1211 Geneva 23, Switzerland}

\author{S. Rigolin}
\email{stefano.rigolin@pd.infn.it}
\affiliation{Dipartimento di Fisica ``G.~Galilei'', Universit\`a di Padova, and\\
INFN, Sezione di Padova, Via Marzolo~8, I-35131 Padua, Italy}

\begin{abstract}
A dynamical origin of the Yukawa couplings is a promising scenario to explain the flavour puzzle. The focus 
of this letter is set on the role of the neutrino Majorana character: when an $O(2)_{N}$ flavour symmetry 
acts on the right-handed neutrino sector, the minimum of the scalar potential allows for large mixing angles 
-in contrast to the simplest quark case- and predicts a maximal Majorana phase. This leads to a strong 
correlation between neutrino mass hierarchy and mixing pattern. Realistic solutions point to the existence of three heavy right-handed neutrinos.
\end{abstract}
\maketitle

%
%

Yukawa couplings are the source of flavour in the Standard Model and describe the large heterogeneity of fermion 
masses and mixings. Understanding the origin of their structure would be tantamount to solving the long-standing flavour 
puzzle.

The concept that quark masses and the Cabibbo angle could arise from extremizing  the possible chiral $SU(3)\otimes SU(3)$ invariant functions was first introduced by N.~Cabibbo in the sixties. Subsequently in Refs.~\cite{Michel:1971th} and \cite{Cabibbo:1960}, group theoretical methods were developed in order to identify their natural extrema.

In 1979, Froggatt and Nielsen~\cite{Froggatt:1978nt} first proposed that Yukawa couplings could correspond to 
dynamical fields, i.e. fields that develop vacuum expectation values in flavour space. Their attempt was 
based on a global $U(1)$ symmetry acting horizontally on the different fermion families and was able to describe 
fermion masses and mixings in agreement with present observations (see Refs.~\cite{Buchmuller:2011tm,
Altarelli:2012ia} for a recent discussion). However, flavour changing neutral currents (FCNCs) in general do 
appear, representing a dangerous drawback for this model.

The Froggatt-Nielsen idea was followed by several proposals based on different type of flavour symmetries 
(i.e continuous or discrete, Abelian or non-Abelian \dots). Among them, a particular role is played by the flavour 
symmetry of the kinetic terms: in the limit of vanishing masses, the fermions of the same type are indistinguishable 
and an $U(3)$ symmetry emerges. Working on this setup, the Minimal Flavour Violation (MFV) ansatz 
\cite{Chivukula:1987py,Hall:1990ac,D'Ambrosio:2002ex,Cirigliano:2005ck,Davidson:2006bd,Kagan:2009bn,Gavela:2009cd,Joshipura:2009gi,
Feldmann:2009dc,Alonso:2011yg,Alonso:2011jd,Alonso:2012fy} has been formulated: it proposes that Yukawa couplings are the only vehicles of flavour at low-energy. Consequently, the Yukawa couplings are the only source of flavour and CP violation in the SM and beyond. A byproduct of this framework is that the energy scale of any New Physics (NP) satisfying the MFV ansatz may be as low as few TeV~\cite{D'Ambrosio:2002ex,Cirigliano:2005ck,Davidson:2006bd,Joshipura:2009gi,Alonso:2011yg,Lalak:2010bk,Fitzpatrick:2007sa,Grinstein:2010ve,Buras:2011wi,Barbieri:2011ci,Alonso:2012jc, Blankenburg:2012nx,Alonso:2012pz,Lopez-Honorez:2013wla}, while in general it should be larger than hundreds of TeV~\cite{Isidori:2010kg}. 

A key aspect of the MFV context is that the Yukawa couplings are promoted to be spurion fields transforming under 
the flavour symmetry, such that the full Lagrangian is formally invariant under the symmetry of the kinetic terms. 
Only once the Yukawa spurions acquire specific background values, fermion masses and mixings can be suitably 
described. It is to be noticed, however, that in the MFV context there is no explanation of the origin of fermion 
masses and mixing, or equivalently there is no explanation for the background values of the Yukawa spurions. 
This motivates the studies performed in Refs.~\cite{Alonso:2011yg,Alonso:2012fy} (see Refs.~\cite{Anselm:1996jm,Barbieri:1999km,Berezhiani:2001mh,Harrison:2005dj} for earlier attempts towards a dynamical origin of the Yukawa couplings and Refs.~\cite{Nardi:2011st,Espinosa:2012uu} for alternative analyses), where the Yukawa spurions are promoted to scalar dynamical fields, invariant under the SM gauge symmetry but transforming under the flavour symmetry. The case with a one-to-one correlation between Yukawa couplings $Y_i$ and dynamical scalar fields $\cY_i$ in the bi-fundamental of the flavour symmetry, $Y_i\equiv \mean{\cY_i}/\Lambda_f$ with $\Lambda_f$ the cutoff of the theory, is discussed at length in Refs.~\cite{Alonso:2011yg,Alonso:2012fy}, but other possibilities, such as $Y_i\equiv \mean{\chi^1_i}\mean{\chi^2_i}/\Lambda^2_f$ with $\chi^{1,2}_i$ scalar fields in the fundamental, have been also considered~\cite{Alonso:2011yg}. The scalar potential constructed out of these fields was studied at the renormalisable level (and adding non-renormalisable terms in the quark context): these effective Lagrangian expansions are possible under the assumption that the ratio of the flavon vevs and the cutoff scale of the theory is smaller than 1, condition that is always satisfied but for the top Yukawa coupling. In this case a non-linear description would be more suitable.

It turns out that the Majorana character of neutrinos can have a deep impact on the nature of the scalar 
potential minima. In Ref.~\cite{Alonso:2012fy}, a particular Type I SeeSaw model with two right-handed 
(RH) neutrinos was analysed: the corresponding flavour symmetry contains an $O(2)_{N}$ factor in the RH 
neutrino sector. At the minimum of the scalar potential, a large mixing angle and a maximal Majorana phase were found. 
This is in contrast to the simplest quark case - which yields a vanishing mixing angle - and it leads to a strong 
correlation between neutrino mass hierarchy and mixing pattern: a novelty in the field (see Refs.~\cite{Altarelli:2010gt,
Ishimori:2010au,Grimus:2011fk,Altarelli:2012bn,Altarelli:2012ss,Bazzocchi:2012st} for recent reviews on lepton flavour models).

The present letter focuses on the lepton sector. We extend the conclusions presented in 
Ref.~\cite{Alonso:2012fy} to generic type I SeeSaw models and explore realistic three-family spectra. The notation 
is  introduced in sect.~\ref{sect:Notation} and then
Sect.~\ref{sect:2Fam} deals with the two-family scenario for the cases: i) 
generic right-handed (RH) neutrino masses; ii) degenerate RH neutrino masses; iii) finally, 
promoting the RH neutrino mass matrix to be as well a dynamical field. Sect.~\ref{sect:3Fam} is devoted 
to the three-family case, identifying the most promising scenario for describing lepton masses and mixing.

\section{MFV in the lepton sector}
\label{sect:Notation}

The MFV ansatz in the lepton sector has been codified in Refs.~\cite{Cirigliano:2005ck,Davidson:2006bd,Gavela:2009cd,Joshipura:2009gi,Alonso:2011jd}. The flavour symmetry of the leptonic kinetic terms, for the type I SeeSaw theory with three RH neutrinos, is 
\beq
G_f=U(3)_{\ell_L}\times U(3)_{E_R}\times U(3)_{N}\,,
\eeq
and the corresponding transformation properties\footnote{Only the transformation properties under the non-Abelian 
part of $G_f$ are explicitly shown.} for leptons are
\beq
\ell_L\sim(3,1,1)\,,\quad
E_R\sim(1,3,1)\,,\quad
N_R\sim(1,1,3)\,.
\eeq
The Yukawa and Majorana mass terms,
\beq
-\LL_Y=\ov{\ell}_LY_EHE_R+\ov{\ell}_LY_\nu\tilde{H}N_R+\ov{N}^c_R\dfrac{M_N}{2}N_R+\hc\,,
\label{Lagrangian}
\eeq 
explicitly break the flavour symmetry, unless the Yukawa couplings and the mass matrix for the RH neutrinos $M_N$ 
are promoted to spurion fields transforming under $G_f$ as 
\beq
Y_E\sim(3,\bar3,1)\,,\quad
Y_\nu\sim(3,1,\bar3)\,,\quad
M_N\sim (1,1,\bar6)\,.
\label{YTransf}
\eeq
Lepton masses and mixings are obtained once these spurions acquire background values. The number of parameters 
described, however, is much larger than the low-energy observables, as expected in the type I SeeSaw context. This 
in general prevents a direct link among neutrino parameters and flavour violating observables. To establish this  
connection the number of spurions has to be reduced from three to two: in Ref.~\cite{Cirigliano:2005ck} $M_N
\propto\unity$; in Ref.~\cite{Gavela:2009cd}, restricting to the two-family RH neutrino case, $M_N\propto\sigma_1$; 
in Ref.~\cite{Alonso:2011jd} $Y^\dag_\nu Y_\nu\propto\unity$. All these cases can be generically described by using 
the Casas-Ibarra parametrization \cite{Casas:2001sr}. In the basis of diagonal mass matrices for RH and left-handed 
(LH) neutrinos and charged leptons, the neutrino Yukawa coupling is written as 
\beq
Y_\nu=\dfrac{1}{v}U\sqrt{\hat m_\nu}R\sqrt{\hat M_N},
\label{YnuCI}
\eeq
where $v$ is the electroweak vev, the hat stands for diagonal mass matrices, $U$ refers to the PMNS matrix and 
$R$ is a complex orthogonal matrix, $R^T R=\unity$. A correct description of lepton masses and mixings is then 
achieved assuming that $Y_E$ acquires the background value:
\beq
Y_E=\cy_E\equiv{\rm diag}(y_e,\,y_\mu,\,y_\tau)\,,
\eeq
while the remaining spurion, $M_N$ or $Y_\nu$ -depending which case is considered- accounts for the neutrino masses 
and the PMNS matrix (see Refs.~\cite{Cirigliano:2005ck,Gavela:2009cd,Alonso:2011jd}). 

To endow the Yukawa couplings with a dynamical context, the spurion fields $Y_E$, $Y_\nu$ and $M_N$ can be promoted to dynamical fields $\cY_E$, $\cY_\nu$ and 
$\cM_N$ respectively. The flavour symmetry is spontaneously broken\footnote{In order to avoid the presence of 
Goldstone bosons, corresponding to the spontaneous breaking of the global flavour symmetry, $G_f$ can be gauged. See 
Refs.~\cite{Grinstein:2010ve,Feldmann:2010yp,Guadagnoli:2011id,Buras:2011zb,Buras:2011wi} for recent developments 
in this direction.} once these fields develop a non-vanishing vev:
\beq
\mean{\cY_E}=\Lambda_f Y_E\,,\quad\,\,
\mean{\cY_\nu}=\Lambda_f Y_\nu\,,\quad\,\,
\mean{\cM_N}= M_N\,.
\eeq
The scale $\Lambda_f$ could be distinct from the scale $M_N$, as the flavour symmetry and the lepton number could be broken at different energies.
The analysis of the scalar potential constructed out of $\cY_E$, $\cY_\nu$ and $\cM_N$ will then establish whether 
realistic masses and mixings correspond to a minimum. A useful earlier analysis of invariants in view of minimising flavour potentials can be found in Refs.~\cite{Jenkins:2007ip,Jenkins:2009dy}.  There are as many independent invariants as independent physical quantities; in this respect see for instance the counting in Ref.~\cite{Broncano:2002rw} for  the type I SeeSaw model including the case of heavy degenerate neutrinos.

\section{The two-family case}
\label{sect:2Fam}

The reduced number of parameters of the two-family scenario allows for a simple analysis. Also, since the tau mass maximally breaks the flavour symmetry, the two-family scenario can be considered an instructive preliminary exercise. 
Generic RH neutrino masses will be first discussed, while the case with degenerate RH neutrinos will follow, for which only two flavons are needed, i.e. $\cY_E$ and $\cY_\nu$. Finally, the RH neutrino mass matrix will also be promoted to a dynamical field. \\

\boldmath
\noindent
$G_f=U(2)_{\ell_L}\times U(2)_{E_R}$\\[2mm]
\unboldmath
This is the flavour symmetry exhibited by the type I SeeSaw Lagrangian in 
the limit of vanishing Yukawa couplings, with generic RH neutrino masses, $M_1\neq M_2$. 
Five independent invariants form a basis\footnote{As 
$G_f$ contains the $U(2)_{\ell_L}$ factor, the operator $\det\left(\cY_E\right)$ is not an invariant, at variance with  
 Ref.~\cite{Alonso:2012fy}, where the analysis of the flavour symmetry concentrated in the case $SU(2)_{\ell_L}$.} at the 
renormalisable level:
\beq
\begin{aligned}
&\tr\left[\cY_E\cY_E^\dagger\right]\,,\qquad
&&\tr\left[\cY_\nu\cA \cY_\nu^\dagger\right]\,,\\
&\tr\left[\left(\cY_E\cY_E^\dagger\right)^2\right]\,, \qquad 
&&\tr\left[\left(\cY_\nu\cA \cY_\nu^\dagger\right)^2\right]\,,\\
&\tr\left[\cY_E\cY_E^\dagger\cY_\nu\cA \cY_\nu^\dagger\right]\,,
\end{aligned}
\label{LeptonInvariants}
\eeq
where $\cA$ is a generic $2\times2$ matrix, here taken to be hermitian in the convention in which the coefficients of the potential are real. The insertion of the matrix $\cA$ is a novelty with respect to the quark case and it is due to the transformation properties of the neutrino Yukawa flavon in this case, $\cY_\nu\sim(2,1)$.

With the set of invariants above, one can construct the corresponding renormalisable scalar potential and 
minimise it. The terms in the first two lines of Eq.~(\ref{LeptonInvariants}) turn out to be responsible for 
fixing the lepton mass hierarchies, while the last term is the only one involving the mixing angle:
\beq
\tr\left[\cY_E\cY_E^\dagger\cY_\nu\cA\cY_\nu^\dagger\right]\propto \,\tr\left[\cy^2_E\, U\sqrt{\hat m_\nu}\, 
P\sqrt{\hat m_\nu}\, U^\dag\right]\,,
\label{LeptMixInv}
\eeq
where  the matrix $P$ encodes the dependence on the high-energy parameters (see Eq.~(\ref{YnuCI})),
\beq
P\equiv R\, \sqrt{\hat M_N}\cA\sqrt{\hat M_N}\, R^\dag\,.
\label{P}
\eeq
 Defining the two-generation PMNS matrix as
\beq
U=\left(
\begin{array}{cc}
 \cos\theta & \sin\theta  \\
 -\sin\theta &  \cos\theta \\
\end{array}
\right)\left(
\begin{array}{cc}
 e^{-i\alpha} &   \\
  & e^{i\alpha}  \\
\end{array}
\right)\,,
\eeq
with $\theta\in[0,\pi/2]$ and $\alpha\in[0,\pi]$, and minimising the invariants in Eq.~(\ref{LeptMixInv}), the following two conditions result:
\begin{align}
&2(y_\mu^2-y_e^2)\sqrt{m_1\,m_2}\sin2\theta\left|P_{12}\right|\sin(2\alpha-\arg P_{12})=0\,,\nn\\
&(y_\mu^2-y_e^2)\Big[\sin2\theta\left(m_1\,P_{11}-m_2\,P_{22}\right)+\\
&\qquad-2\cos2\theta\sqrt{m_1\,m_2}\left|P_{12}\right|\cos\left(2\alpha-\arg P_{12}\right)\Big]=0\,,\nn
\end{align}
where $m_{1,2}$ are the eigenvalues of $\hat m_\nu$. For non-trivial mixing ($\sin 2\theta\ne 0$) and neglecting the trivial solutions (i.e. degenerate charged lepton masses, 
vanishing neutrino masses, or vanishing $|P_{12}|$),  it follows that 
\beq
\begin{aligned}
&2\alpha-\arg P_{12}=n\pi\,,\qquad\text{with}\,\,\,\, n\in  \mathbb{Z}\\
&\tan2\theta=2\left|P_{12}\right|\dfrac{\sqrt{m_1\,m_2}}{m_1\,P_{11}-m_2\,P_{22}}\,\cos(2\alpha-\text{arg}P_{12})\,.
\end{aligned}
\label{LeptonSolutions}
\eeq
The first expression connects the low-energy and the high-energy phases, while the second one represents a link among 
the size of the mixing angle and the type of  neutrino spectrum. It is  precisely the Majorana neutrino character that allows 
this novel connection, through the solutions with non-trivial Majorana CP phases. However, the presence of the generic matrix $\cA$  in Eq.~(\ref{P}) prevents clear predictions for the mixing angle.\\[2mm]

\boldmath
\noindent
$G_f=U(2)_{\ell_L}\times U(2)_{E_R}\times O(2)_{N}$\\[2mm]
\unboldmath
This case corresponds to the degenerate RH neutrino masses, $M_1=M_2\equiv M$. This is the largest possible symmetry in the RH neutrino sector, once non-vanishing masses for the heavy RH neutrinos are considered. The invariants that form a basis at the renormalisable level are those in Eq.~(\ref{LeptonInvariants}), but since now $\cY_\nu
\sim(2,1,\bar 2)$, $\cA$ can only take the values $\cA=\unity$ and $\cA=\sigma_2$. The latter 
 leads to only one non-trivial invariant,
\beq
\tr\left[\left(\cY_\nu\sigma_2\cY_\nu^\dagger\right)^2\right]\,,
\eeq
which can be rewritten as
\beq
\tr\left[\cY_\nu\cY_\nu^T\cY_\nu^*\cY_\nu^\dag\right]\,.
\label{O2Invariant}
\eeq
In summary, the basis of invariants in this case is constituted by those in Eq.~(\ref{LeptonInvariants}) with $\cA=\unity$, plus the operator in Eq.~(\ref{O2Invariant}).

In the present case of degenerate heavy neutrinos, the $R$ (and thus $P$) matrix simplifies to:
\beq
\begin{aligned}
R&=\left(
\begin{array}{cc}
\cosh\omega  & -i\sinh\omega \\
i\sinh\omega  & \cosh\omega \\
\end{array}
\right)\,,\\
P&=M\,\left(
\begin{array}{cc}
\cosh2\omega  & -i\sinh2\omega \\
i\sinh2\omega  & \cosh2\omega \\
\end{array}
\right)\,.
\end{aligned}
\eeq
The conditions which define the minima, Eq.~(\ref{LeptonSolutions}), become in turn
\beq
\begin{aligned}
&\alpha=\pi/4\qquad \text{or}\qquad \alpha=3\pi/4\,,\\
&\tan2\theta=2\sin2\alpha\dfrac{\sqrt{m_1\,m_2}}{m_1-m_2}\tan2\omega\,.
\end{aligned}
\label{LeptonSolutionsDegenerate2}
\eeq
A maximal relative Majorana phase is thus obtained; it does not imply experimental consequences for CP-odd observables, though,  as the relative Majorana phase among the two neutrino eigenvalues is $\pi/2$.  Eq.~(\ref{LeptonSolutionsDegenerate2}) defines a class of extrema of the scalar potential: in particular, a large mixing 
angle is obtained for almost degenerate masses, while a small angle follows in the hierarchical case. It is then necessary 
to discuss the full minimisation of the scalar potential to identify the configuration of angles corresponding to the absolute 
minimum. Eq.~(\ref{LeptonSolutionsDegenerate2}) agrees with the results in Ref.~\cite{Alonso:2012fy}, for a particular choice of $\omega$.\footnote{In the notation of Ref.~\cite{Alonso:2012fy}, $\omega$ is defined as $e^\omega\equiv\sqrt{y/y'}$.} As shown there, degenerate neutrino masses are a minimum of the scalar potential and 
therefore the maximal angle solution is also a minimum.

Note that the results in Eq.~(\ref{LeptonSolutionsDegenerate2}) stem from the last invariant in Eq.~(\ref{LeptonInvariants})  with $\cA=\unity$, and in particular are not affected 
by the introduction of the new invariant in Eq.~(\ref{O2Invariant}). The latter has an impact only on the neutrino  
spectrum and fixes $\omega$:  for $\omega=0$  the mixing angle vanishes, see  Eq.~(\ref{LeptonSolutionsDegenerate2}) (and Ref.~\cite{Alonso:2012fy} for more details);  for $\omega\ne0$
 it allows instead a degenerate mass configuration at the minimum and therefore selects the configuration with  
maximal angle. 

It is interesting to recover the minima of the scalar potential using a different parametrisation than the Casas-Ibarra one: 
the bi-unitary parametrisation. The latter consists in decomposing a matrix as a product of a unitary matrix, a diagonal matrix 
of eigenvalues and a second unitary matrix. Without loss of generality, we will work in the basis in which the RH neutrino and the charged lepton mass matrices 
are diagonal. The neutrino Yukawa coupling (vev of the $\cY_\nu$ field) can be written as 
\beq
Y_\nu\equiv U_L \hat Y_\nu U_R
\label{BiUnitaryNotation}
\eeq
with $U_{L,R}$ being unitary matrices and $\hat Y_\nu={\rm diag}(y_{\nu_1},\,y_{\nu_2})$. The light neutrino mass matrix reads then 
\beq
m_\nu=v^2\,Y_\nu\dfrac{1}{M_N} Y^T_\nu=v^2\, U_L \hat Y_\nu U_R \dfrac{1}{M_N}U^T_R \hat Y_\nu U^T_L\,.
\eeq
Using Von Neumann's trace inequality and the freedom to redefine the electron and  muon fields,  the analysis 
of $\tr\left[\cY_E\cY_E^\dagger\cY_\nu\cY_\nu^\dagger\right]$ leads immediately to
\beq
U_{L}\propto
\left(
\begin{array}{cc}
 1 & 0 \\
 0 & 1 \\
\end{array}
\right)\,,
\label{LeptonSolutionBiUnitary1}
\eeq
where unphysical phases have been dropped for simplicity. On the other side, the invariant in Eq.~(\ref{O2Invariant}) leads to the following structure for $U_R$ at the potential minimum:
\beq
U_{R}U^T_{R}\propto\left(
\begin{array}{cc}
 0 & 1 \\
 1 & 0 \\
\end{array}
\right)\,,
\label{LeptonSolutionBiUnitary2}
\eeq 
besides the trivial one. From Eqs.~(\ref{LeptonSolutionBiUnitary1}) and (\ref{LeptonSolutionBiUnitary2}), the light neutrino mass matrix takes the form
\beq
\hat m_\nu=U^T m_\nu U=\dfrac{v^2}{M}\,\tilde y_{\nu_1} \tilde y_{\nu_2}\,U^T\left(
\begin{array}{cc}
 0 & 1 \\
 1 & 0 \\
\end{array}
\right)U\,,
\label{NeutrinoMassO2}
\eeq
where  the unphysical phases eventually present in $U_{L,R}$ have been reabsorbed in $\tilde y_{\nu_i}$. As a result, the PMNS matrix reads:
\beq
U=\left(
\begin{array}{cc}
\sqrt2/2  & \sqrt2/2 \\
-\sqrt2/2  & \sqrt2/2 \\
\end{array}
\right)
\left(
\begin{array}{cc}
 i &  \\
  & 1 \\
\end{array}
\right)\,.
\eeq
In conclusion, the general SeeSaw setup with heavy degenerate neutrinos provides a solution with: i) a degenerate light neutrino spectrum; ii) a correlated maximal mixing angle $\theta=\pi/4$; iii)  and a correlated maximal relative Majorana phase $2\alpha=\pi/2$. This spectrum is  in agreement 
with the analytical results in Eq.~(\ref{LeptonSolutionsDegenerate2}) and subsequent discussion. \\[2mm]

\boldmath
\noindent
$G_f=U(2)_{\ell_L}\times U(2)_{E_R}\times U(2)_{N}$\\[2mm]
\unboldmath
A pertinent question is whether other global flavour symmetries may produce the same or similar results than those found above. In the case in which the RH neutrino sector exhibits a $U(2)_{N}$ factor, the invariance of the complete Lagrangian under $G_f$ requires $M_N$ to be also promoted to a dynamical field,  properly transforming 
under $U(2)_{N}$. The operators in Eq.(\ref{LeptonInvariants}) are then invariants of $G_f$ only for $\cA=\unity$, while  Eq.~(\ref{O2Invariant}) is not an invariant. On the other side, another three additional 
operators are allowed at the renormalisable level and enlarge the operator basis:
\beq
\begin{gathered}
\tr\left[\cM_N^*\cM_N\right]\,,\qquad
\tr\left[\left(\cM_N^*\cM_N\right)^2\right]\,,\\ 
\tr\left[\cM_N^*\cM_N\cY_\nu^\dagger\cY_\nu\right]\,.
\end{gathered}
\label{LeptonInvariantsSU}
\eeq
The last invariant in this list leads to $U_{R}\propto\unity$, or equivalent configurations, as it is straightforward to prove using the bi-unitary parametrisation in Eq.~(\ref{BiUnitaryNotation}) and  Von Neumann's trace inequality. Together with Eq.~(\ref{LeptonSolutionBiUnitary1})  for  $U_{L}$,  the minimum would indicate  a vanishing mixing angle~\footnote{More precisely, $\sin\theta=0$: the configurations with angle $\pi/2$ lead to no mixing after ``reordering" the mass states.}.

Summarizing: in the two-family case, only when the flavour symmetry of the type I SeeSaw encodes  $O(2)_{N}$  a maximal mixing angle can be predicted at the minimum of the scalar potential, together with a relative Majorana phase of $\pi/2$ and a degenerate light neutrino spectrum.

Moreover, the previous discussion shows that, when Yukawas have a dynamical origin, a connection between low-energy and high-energy parameters is possible. For instance, the CP-asymmetry entering Leptogenesis for the case of $G_f=U(2)_{\ell_L}\times U(2)_{E_R}\times O(2)_{N}$ is a function of $\text{arg}(P_{12})$ (see the phase relation in Eq.~(\ref{LeptonSolutions})). Would that relation hold exactly in nature, the maximal relative CP phase obtained would entail no leptogenesis; nevertheless, departures from that precise relation are to be expected in a realistic scenario, and they may suffice as seeds of the matter-antimatter asymmetry.  This subject deserves further future exploration.

\section{The three-family case}
\label{sect:3Fam}

For the realistic scenario with three lepton families,  it is interesting to analyse both the cases with three  
 and  with  two heavy RH neutrinos (as experimentally one of the light neutrino is allowed to be massless).
\subsection{Two RH neutrinos}
\label{2RHcase}

Generic RH neutrino masses or the consideration of $M_N$ as a flavon would not lead to any improvement with respect to the two-family case. For the interesting scenario with degenerate RH neutrino masses, i.e. $G_f$ containing the factor $O(2)_{N}$, 
as the large angle corresponds always to the most degenerate sector, it suggests to identify it with the ``solar" angle $\theta_{12}$: an  explicity analysis shows  it to lie then in the wrong quadrant, 
while furthermore no other angles arise (see Ref.~\cite{Alonso:2012fy} for further details in a concrete example).  Therefore, this avenue does not lead to a realistic pattern, at least in its simplest formulation.

\boldmath
\subsection{Three RH neutrinos}
\unboldmath

Most of the results in the previous section can be  
generalised to the case of three RH neutrinos.  \\[2mm]

\boldmath
\noindent
$G_f=U(3)_{\ell_L}\times U(3)_{E_R}$\\[2mm]
\unboldmath
This symmetry 
holds when neither is $M_N$  promoted to be a dynamical field nor a degeneracy among its eigenvalues is present.
The invariants defining a basis at the renormalisable level are still those listed in Eq.~(\ref{LeptonInvariants}), 
 with the obvious three-dimensional generalisation of the matrices $Y_i$ and $\cA$. The minimisation procedure turns out to be more complicated technically than  that leading to Eq.~(\ref{LeptonSolutions}). In  any case,  the presence of the generic matrix $\cA$ prevents also in this case to make clear predictions for the mixing angles.\\[2mm]

\boldmath
\noindent
$G_f=U(3)_{\ell_L}\times U(3)_{E_R}\times U(3)_{N}$\\[2mm]
\unboldmath
In this case, promoting  $M_N$ to be a flavon field transforming under the $U(3)_{N}$ subgroup of the flavour symmetry, the possible invariants are the three-family equivalent to those in Eqs.~(\ref{LeptonInvariants}) with $\cA=\unity$, plus those in Eq.~(\ref{LeptonInvariantsSU}). The bi-unitary parametrisation for $Y_\nu$ is a useful tool also in this case:  it is straightforward to verify that, at the minimum of the potential, both $U_L$ and $U_R$ are proportional to the identity matrix (or equivalent configurations, after redefining the fermion fields), and in consequence lead to no mixing.

The results presented in the previous sections are strictly valid considering the scalar potential at the renormalisable 
level, as higher order operators may be neglected under the assumption that the ratio of the flavon vevs and 
the cutoff of the theory is smaller than one. Adding non-renormalisable terms to the lepton scalar potential is currently 
under investigation. 

Moreover, it has been implicitly assumed above  that only fields transforming 
in the bi-fundamental representation of the flavour symmetry could be used when constructing the invariant operators. An interesting possibility, already analysed for the quark case~\cite{Alonso:2011yg}, is to introduce in addition fields in 
the fundamental representation of $G_f$ and analyse the interplay of both type of fields. This can naturally 
happen when two, out of three, RH neutrinos are degenerate in mass, as we are going to consider in the following.\\[2mm]

\boldmath
\noindent
$G_f=U(3)_{\ell_L}\times U(3)_{E_R}\times O(2)_{N}$\\[2mm]
\unboldmath
Let us consider this symmetry when two, out of three, RH neutrinos are degenerate in mass. One could think that this setup is equivalent to that with only two RH neutrinos, degenerate in mass, previously discussed 
in sect.~\ref{2RHcase}, as the flavour symmetry is the same. Nevertheless, the presence of a third non-degenerate state increases the number of invariants that can be built. We will show next that, due to the interplay between the doublet and the singlet states, all three mixing angles can be non-vanishing.

The leptonic flavour Lagrangian is given in this case by
\beq
\begin{split}
-\LL_Y=&\ov{\ell}_LY_EHE_R+\ov{\ell}_LY'_\nu\tilde{H}N'_R+\ov{\ell}_LY_\nu\tilde{H}N_R+\\
&+\dfrac{M'}{2}\ov{N}^{\prime c}_RN'_R+\dfrac{M}{2}\ov{N}^c_R\unity N_R+\hc\,,
\end{split}
\label{Lagrangian2}
\eeq 
where $N_R$ ($N'_R$) is a doublet (singlet) of $O(2)_{N}$, and the flavons associated to the neutrino Yukawa couplings are
\beq
\cY_\nu\sim(3,1,\bar2)\,,\qquad
\cY'_\nu\sim(3,1,1)\,.
\eeq
Once the Yukawa flavons develop vevs, the light neutrino mass matrix is generated: 
\beq
m_\nu=\dfrac{v^2}{M'}Y'_\nu Y^{\prime T}_\nu+\dfrac{v^2}{M}Y_\nu Y^T_\nu\,.
\eeq

A total of nine independent invariants at the renormalisable level can be constructed in this case, namely
\beq
\begin{gathered}
\tr\left[\cY_E\cY_E^\dagger\right]\,,\qquad
\tr\left[\cY_\nu\cY_\nu^\dagger\right]\,,\qquad
\cY^{\prime \dag}_\nu \cY'_\nu\,,\\
\tr\left[\left(\cY_E\cY_E^\dagger\right)^2\right]\,, \qquad 
\tr\left[\left(\cY_\nu\cY_\nu^\dagger\right)^2\right]\,,\\
\tr\left[\cY_E\cY_E^\dagger\cY_\nu\cY_\nu^\dagger\right]\,, \qquad 
\tr\left[\cY_\nu\cY_\nu^T\cY_\nu^*\cY_\nu^\dag\right]\,,\\
\cY^{\prime \dag}_\nu \cY_E \cY^\dag_E \cY'_\nu \,,\qquad
\cY^{\prime \dag}_\nu \cY_\nu \cY^\dag_\nu \cY'_\nu \,.
\end{gathered}
\eeq
The corresponding renormalisable scalar potential can be written as a sum of three different terms:
\beq
V=V_\Delta+V_L+V_R\,,
\eeq
with
\beq
\begin{aligned}
V_\Delta&=-(\mu_E^2, \mu_\nu^2, \mu^{\prime 2}) \cX^2 + \cX^{2\dag}\lambda \cX^2+\\
&\qquad\qquad+\lambda_E\,\tr\left[\left(\cY_E\cY_E^\dagger\right)^2\right]+\lambda_\nu\,\tr\left[\left(\cY_\nu\cY_\nu^\dagger\right)^2\right]\,,\\
V_L&=g_a \tr\left[\cY_E\cY_E^\dagger\cY_\nu\cY_\nu^\dagger\right] + 
g_b \cY^{\prime \dag}_\nu \cY_E \cY^\dag_E \cY'_\nu +
g_c \cY^{\prime \dag}_\nu \cY_\nu \cY^\dag_\nu \cY'_\nu\,,\\
V_R&=h' \tr\left[\cY_\nu\cY_\nu^T\cY_\nu^*\cY_\nu^\dag\right]\,,
\end{aligned}
\label{V3terms}
\eeq
where $\cX^2\equiv\left(\tr\left[\cY_E\cY_E^\dagger\right],\tr\left[\cY_\nu\cY_\nu^\dagger\right],\cY^{\prime \dag}_\nu\cY'_\nu\right)^T$,  
and $\lambda$ is a $3 \times 3$ matrix of quartic couplings. The minimisation of the scalar potential will be implemented using the bi-unitary parametrisation for the vevs of the neutrino Yukawa flavons,
\beq
Y_\nu\equiv U_L \left(
\begin{array}{cc}
0  & 0  \\
y_{\nu_1}  & 0  \\
0  & y_{\nu_2}  \\
\end{array}
\right)
 U_R\,,\qquad
Y'_\nu\equiv U'_L y'_\nu\,,
\label{BiUnitaryParGood}
\eeq
where $U_{L,R}$ are $3\times3$ and $2\times2$ unitary matrices, respectively, and $U'_L$ a unitary vector in the 
$U(3)_{\ell_L}$ space~\footnote{ 
Other two configurations for $Y_\nu$ are possible, permuting the rows in Eq.~(\ref{BiUnitaryParGood});  similar conclusions are obtained with them.}.

The dependence on the physical parameters of the three terms in the scalar potential is as follows: i) $V_\Delta$ depends only on the 
eigenvalues of $Y_E$, $Y_\nu$ and $Y'_\nu$; ii) $V_L$ depends on those eigenvalues  and on $U_L$ and $U'_L$; iii) finally, $V_R$ depends 
only on $U_R$ and the eigenvalues. As a result, $V_R$ is minimized when $U_R$ takes the values 
\beq
\begin{aligned}
&U_{R}U^T_{R}\propto\unity\,,
\qquad&&\text{for}\qquad h'<0\,,\\
&U_{R}U^T_{R}\propto
\left(
\begin{array}{cc}
 0 & 1 \\
 1 & 0 \\
\end{array}
\right)\,,
\qquad&&\text{for}\qquad h'>0\,.
\end{aligned}
\eeq 
The minimisation of $V_L$ is cumbersome as three terms contribute and the absolute minimum is determined by the relative signs  of those three terms, which in turn  
 depend on $U_L$, $U'_L$ and on the product 
$U^{\prime\dag}_L U_L$, respectively.  The possible configurations that minimise each of the terms are then
\begin{align}
&\begin{cases}
U_L\propto\unity & \qquad g_a<0\\
U_L\propto\left(
\begin{array}{ccc}
 0 & 0 & 1 \\
 0 & 1 & 0 \\
 1 & 0 & 0 \\
\end{array}
\right)
 & \qquad g_a>0\\
\end{cases}\\
&\begin{cases}
U^{\prime\dag}_L\propto \left(
\begin{array}{ccc}
 0 & 0 & 1 \\
\end{array}
\right) & \qquad g_b<0\\[1mm]
U^{\prime\dag}_L\propto
\left(
\begin{array}{ccc}
 1 & 0 & 0 \\
\end{array}
\right)
 & \qquad g_b>0\\
\end{cases}\\
&\begin{cases}
U^{\prime\dag}_LU_L\propto \left(
\begin{array}{ccc}
 0 & 0 & 1\\
\end{array}
\right) & \qquad g_c<0\\[1mm]
U^{\prime\dag}_LU_L\propto
\left(
\begin{array}{ccc}
 1 & 0 & 0 \\
\end{array}
\right)
 & \qquad g_c>0\,.
\end{cases}
\end{align} 
In consequence, when considering the full minimisation of $V_L$, there are four cases in which all the three terms select the same vacuum and a precise prediction for the light neutrino mass matrix can follow: when the product of $g_a$, $g_b$ and $g_c$ is negative. Defining for compactness $z\equiv y_{\nu_1} y_{\nu_2} v^2/ M$ and $z'\equiv y^{\prime 2}_\nu v^2/ M'$, 
the four cases are
\begin{enumerate}
\item $g_a>0,g_b>0,g_c<0$:
\beq
m_\nu=\left(
\begin{array}{ccc}
z'  & z & 0 \\
z  & 0 & 0 \\
0  & 0 & 0 \\
\end{array}
\right)
\rightarrow
\begin{cases}
\tan2\theta_{12}=z/z'\\
m_{\nu_1}\neq m_{\nu_2}\\
m_{\nu_3}=0
\end{cases}
\label{Sol1}
\eeq
\item $g_a>0,g_b<0,g_c>0$:
\beq
m_\nu=\left(
\begin{array}{ccc}
0 & z & 0 \\
z  & 0 & 0 \\
0  & 0 & z' \\
\end{array}
\right)
\rightarrow
\begin{cases}
\theta_{12}=\pi/4\\
m_{\nu_1}= m_{\nu_2}\neq m_{\nu_3}
\end{cases}
\label{Sol2}
\eeq
\item $g_a<0,g_b>0,g_c>0$:
\beq
m_\nu=\left(
\begin{array}{ccc}
z' & 0 & 0 \\
0  & 0 & z \\
0  & z & 0 \\
\end{array}
\right)
\rightarrow
\begin{cases}
\theta_{23}=\pi/4\\
m_{\nu_1}\neq m_{\nu_2} = m_{\nu_3}
\end{cases}
\label{Sol3}
\eeq
\item $g_a<0,g_b<0,g_c<0$:
\beq
m_\nu=\left(
\begin{array}{ccc}
0  & 0 & 0 \\
0  & 0 & z \\
0  & z & z' \\
\end{array}
\right)
\rightarrow
\begin{cases}
\tan2\theta_{23}=z/z'\\
m_{\nu_2}\neq m_{\nu_3}\\
m_{\nu_1}=0
\end{cases}
\label{Sol4}
\eeq
\end{enumerate}
Case 1 (4) describes an inverse (direct) hierarchical spectrum and only one sizable mixing angle, the solar (atmospheric) one. In case 2, the light neutrinos ${\nu_1}$ and ${\nu_2}$ are degenerate  and both mass orderings (hierarchical or degenerate) can be accommodated, while a maximal solar angle is predicted. Finally, case 3 corresponds to degenerate $\nu_2$ and $\nu_3$: a realistic scenario points to three almost degenerate neutrinos. Note that cases 2 and 3 encompass two degenerate neutrinos and  the relative Majorana phase between the two degenerate states is $\pi/2$. 

Cases 1-4 only account for one sizable angle. Configurations with three non-trivial angles, however, follow in a straightforward way when the product of $g_a$, $g_b$ and $g_c$ in Eq.~(\ref{V3terms}) is positive: the distinct terms in $V_L$ compete then and generic $U_L$ and $U'_L$ are selected at the minimum. These realistic configurations can be thought of as interpolations between the cases in Eqs.~(\ref{Sol1})-(\ref{Sol4}); however, they do not admit a perturbative expansion in the  coefficients $g_a$, $g_b$ and $g_c$, and no simple analytical formulae follow. The setup appears very promising, though, as all three angles can be naturally non-vanishing and moreover the number of free parameters is smaller than the number of observables, leading to predictive scenarios in which mixing angles and Majorana phases are linked to the spectrum. This case is currently under exploration.

\section{Conclusions}

In this letter, the dynamical origin of the Yukawa structure in the lepton sector is investigated in the context of 
type I SeeSaw. Yukawa couplings are promoted to be scalar fields, transforming only under global flavour symmetries, 
and acquiring vevs that minimise the corresponding scalar potential. 

The flavour symmetries that have been considered are the maximal symmetries of the  Lagrangian in the limit of 
vanishing Yukawa couplings and i) generic RH neutrino masses, ii) degenerate RH neutrino masses, iii) promoting to a field the RH neutrino mass matrix itself. 

The relatively large tau lepton mass represents a maximal breaking of the flavour symmetry, and this suggests to consider first the two-family 
scenario. The flavour symmetry for the three scenarios mentioned above is $U(2)_{\ell_L}\times U(2)_{E_R}$,  $U(2)_{\ell_L}\times U(2)_{E_R}
\times O(2)_{N}$ and  $U(2)_{\ell_L}\times U(2)_{E_R}\times U(2)_{N}$, respectively. 
Tantalizingly suggestive results follow when  the flavour group $G_f$ contains  $O(2)_{N}$,  corresponding to degenerate RH neutrino masses. A specific 
model of this type has been previously studied in Ref.~\cite{Alonso:2012fy}, while in this letter we have extended that analysis to generic type I Seesaw scenarios containing two degenerate RH neutrinos. We have proven that the minimum of the potential allows a maximal mixing angle and a maximal Majorana phase, correlated with a degenerate light neutrino spectrum. It is the Majorana neutrino character -technically via the non-trivial Majorana phases- that allows this novel connection.

For  three light generations with only two RH neutrinos, no satisfactory or promising scenario is obtained: even if  a maximal mixing angle is allowed within the most degenerate light sector -the solar one,  it would lie in a quadrant which is experimentally excluded; moreover no other angles appear at this level. On the other hand, when three RH neutrinos are considered with two of them degenerate in mass, the flavour symmetry  $G_f=U(3)_{\ell_L}\times U(3)_{E_R}\times O(2)_{N}$ may lead to realistic patterns of lepton masses and mixings.  Three sizable mixing angles can arise and are determined in terms of lepton and neutrino masses, from the interplay of two different types of Yukawa fields: a field transforming under the bi-fundamental of $G_f$ and an other one under the fundamental. 
In summary, our results indicate that a realistic solution for the Flavour Puzzle in the lepton sector requires three RH neutrinos, two of which must be degenerate. All three light neutrinos would therefore acquire masses, and the precise values of the mixing angles and Majorana phases are related to the specific light mass spectrum.

The analysis illustrated in this letter considered only renormalisable scalar potentials, while the impact of 
non-renormalisable terms and perturbations is currently under investigation.

\section{Note added in proof}

When all three RH neutrinos are degenerate, the flavour symmetry is
\beq
G_f=U(3)_{\ell_L}\times U(3)_{E_R} \times O(3)_{N}\,,
\eeq
and the basis of invariants is composed of the operators in Eq.~(\ref{LeptonInvariants}) with $\cA=\unity$, plus that in  Eq.~(\ref{O2Invariant}). The study of the extrema of these invariants has been recently presented in Ref.~\cite{Alonso:2013nca}. Minimising the corresponding scalar potential as illustrated above, the solutions are consistent with those  in Ref.~\cite{Alonso:2013nca}.

Using Von Neumann's trace inequality and the freedom to redefine the charged lepton fields, $U_L\propto\unity$ at the minimum of the potential, while the product $U_{R}U^T_R$ acquires two possible structures:
\beq
U_{R}U^T_R\propto\unity
\quad\text{or}\quad
U_{R}U^T_R\propto
\left(
\begin{array}{ccc}
 0 & 0 & 1 \\
 0 & 1 & 0 \\
 1 & 0 & 0 \\
\end{array}
\right)\,.
\eeq
While the first solution leads to no mixing, the second one corresponds to a neutrino mass matrix of the type
\beq
\hat m_\nu=\dfrac{v^2}{M}\,U^T\left(
\begin{array}{ccc}
 0 & 0 & \tilde y_{\nu_1} \tilde y_{\nu_3} \\
 0 & \tilde y_{\nu_2}^2 & 0 \\
 \tilde y_{\nu_1}\tilde y_{\nu_3} & 0 & 0 \\
\end{array}
\right)U\,,
\label{NeutrinoMassO3}
\eeq
where $\tilde y_{\nu_i}$ contain the three entries of $\hat Y_\nu$ and unphysical phases. In the normal or inverse hierarchical case, two of the light neutrinos are degenerate in mass and a maximal angle and a maximal Majorana phase arise in their corresponding sector. On the other hand, if the third light neutrino is almost degenerate with the other two, then the perturbations split the spectrum and a second sizable angle arises~\cite{Alonso:2013nca}.

\section*{Acknowledgments}
We thank Gino Isidori and Luciano Maiani for interesting discussions and comments on the preliminary version of this letter. We also acknowledge the lively environment and stimulating discussions with our colleagues of the CERN Theory division.
We acknowledge partial support by European Union FP7 ITN INVISIBLES (Marie Curie Actions, PITN-GA-2011-289442), 
CiCYT through the project FPA2009-09017, CAM through the project HEPHACOS P-ESP-00346, MICINN through the grant BES-2010-037869 and the Juan de la Cierva programme (JCI-2011-09244), Italian Ministero dell'Uni\-ver\-si\-t\`a e della Ricerca Scientifica through the COFIN program (PRIN 2008) and the contracts MRTN-CT-2006-035505 and  PITN-GA-2009-237920 (UNILHC).

\providecommand{\href}[2]{#2}\begingroup\raggedright\endgroup


\begin{thebibliography}{10}

\bibitem{Michel:1971th}
L.~Michel and L.~Radicati,  Proc. of the Fifth Coral Gables Conference on
  Symmetry principles at High Energy, ed. by B. Kursunoglu et al., W. H.
  Benjamin, Inc. New York (1965); Annals Phys. 66 (1971) 758.

\bibitem{Cabibbo:1960}
N.~Cabibbo and L.~Maiani in {\it Evolution of particle physics}, Academic Press
  (1970), 50, App. I.

\bibitem{Froggatt:1978nt}
C.~Froggatt and H.~B. Nielsen,  Nucl.Phys. {\bf B147} (1979) 277.

\bibitem{Buchmuller:2011tm}
W.~Buchmuller, V.~Domcke, and K.~Schmitz,  JHEP {\bf 1203} (2012) 008,
  [\href{http://xxx.lanl.gov/abs/1111.3872}{{\tt arXiv:1111.3872}}].

\bibitem{Altarelli:2012ia}
G.~Altarelli, F.~Feruglio, I.~Masina, and L.~Merlo,  JHEP {\bf 1211} (2012)
  139, [\href{http://xxx.lanl.gov/abs/1207.0587}{{\tt arXiv:1207.0587}}].

\bibitem{Chivukula:1987py}
R.~S. Chivukula and H.~Georgi,  Phys.Lett. {\bf B188} (1987) 99.

\bibitem{Hall:1990ac}
L.~J. Hall and L.~Randall,  Phys. Rev. Lett. {\bf 65} (1990) 2939--2942.

\bibitem{D'Ambrosio:2002ex}
G.~D'Ambrosio, G.~Giudice, G.~Isidori, and A.~Strumia,  Nucl.Phys. {\bf B645}
  (2002) 155--187, [\href{http://xxx.lanl.gov/abs/hep-ph/0207036}{{\tt
  hep-ph/0207036}}].

\bibitem{Cirigliano:2005ck}
V.~Cirigliano, B.~Grinstein, G.~Isidori, and M.~B. Wise,  Nucl.Phys. {\bf B728}
  (2005) 121--134, [\href{http://xxx.lanl.gov/abs/hep-ph/0507001}{{\tt
  hep-ph/0507001}}].

\bibitem{Davidson:2006bd}
S.~Davidson and F.~Palorini,  Phys.Lett. {\bf B642} (2006) 72--80,
  [\href{http://xxx.lanl.gov/abs/hep-ph/0607329}{{\tt hep-ph/0607329}}].

\bibitem{Kagan:2009bn}
A.~L. Kagan, G.~Perez, T.~Volansky, and J.~Zupan,  Phys. Rev. {\bf D80} (2009)
  076002, [\href{http://xxx.lanl.gov/abs/0903.1794}{{\tt arXiv:0903.1794}}].

\bibitem{Gavela:2009cd}
M.~Gavela, T.~Hambye, D.~Hernandez, and P.~Hernandez,  JHEP {\bf 0909} (2009)
  038, [\href{http://xxx.lanl.gov/abs/0906.1461}{{\tt arXiv:0906.1461}}].

\bibitem{Joshipura:2009gi}
A.~S. Joshipura, K.~M. Patel, and S.~K. Vempati,  Phys.Lett. {\bf B690} (2010)
  289--295, [\href{http://xxx.lanl.gov/abs/0911.5618}{{\tt arXiv:0911.5618}}].

\bibitem{Feldmann:2009dc}
T.~Feldmann, M.~Jung, and T.~Mannel,  Phys.Rev. {\bf D80} (2009) 033003,
  [\href{http://xxx.lanl.gov/abs/0906.1523}{{\tt arXiv:0906.1523}}].

\bibitem{Alonso:2011yg}
R.~Alonso, M.~Gavela, L.~Merlo, and S.~Rigolin,  JHEP {\bf 1107} (2011) 012,
  [\href{http://xxx.lanl.gov/abs/1103.2915}{{\tt arXiv:1103.2915}}].

\bibitem{Alonso:2011jd}
R.~Alonso, G.~Isidori, L.~Merlo, L.~A. Munoz, and E.~Nardi,  JHEP {\bf 1106}
  (2011) 037, [\href{http://xxx.lanl.gov/abs/1103.5461}{{\tt
  arXiv:1103.5461}}].

\bibitem{Alonso:2012fy}
R.~Alonso, M.~Gavela, D.~Hernandez, and L.~Merlo,  Phys.Lett. {\bf B715} (2012)
  194--198, [\href{http://xxx.lanl.gov/abs/1206.3167}{{\tt arXiv:1206.3167}}].

\bibitem{Lalak:2010bk}
Z.~Lalak, S.~Pokorski, and G.~G. Ross,  JHEP {\bf 1008} (2010) 129,
  [\href{http://xxx.lanl.gov/abs/1006.2375}{{\tt arXiv:1006.2375}}].

\bibitem{Fitzpatrick:2007sa}
A.~L. Fitzpatrick, G.~Perez, and L.~Randall,  Phys.Rev.Lett. {\bf 100} (2008)
  171604, [\href{http://xxx.lanl.gov/abs/0710.1869}{{\tt arXiv:0710.1869}}].

\bibitem{Grinstein:2010ve}
B.~Grinstein, M.~Redi, and G.~Villadoro,  JHEP {\bf 1011} (2010) 067,
  [\href{http://xxx.lanl.gov/abs/1009.2049}{{\tt arXiv:1009.2049}}].

\bibitem{Buras:2011wi}
A.~J. Buras, M.~V. Carlucci, L.~Merlo, and E.~Stamou,  JHEP {\bf 1203} (2012)
  088, [\href{http://xxx.lanl.gov/abs/1112.4477}{{\tt arXiv:1112.4477}}].

\bibitem{Barbieri:2011ci}
R.~Barbieri, G.~Isidori, J.~Jones-Perez, P.~Lodone, and D.~M. Straub,  Eur.
  Phys. J. {\bf C71} (2011) 1725,
  [\href{http://xxx.lanl.gov/abs/1105.2296}{{\tt arXiv:1105.2296}}].

\bibitem{Alonso:2012jc}
R.~Alonso, M.~Gavela, L.~Merlo, S.~Rigolin, and J.~Yepes,  JHEP {\bf 1206}
  (2012) 076, [\href{http://xxx.lanl.gov/abs/1201.1511}{{\tt
  arXiv:1201.1511}}].

\bibitem{Blankenburg:2012nx}
G.~Blankenburg, G.~Isidori, and J.~Jones-Perez,  Eur.Phys.J. {\bf C72} (2012)
  2126, [\href{http://xxx.lanl.gov/abs/1204.0688}{{\tt arXiv:1204.0688}}].

\bibitem{Alonso:2012pz}
R.~Alonso, M.~Gavela, L.~Merlo, S.~Rigolin, and J.~Yepes,  Phys.Rev. {\bf D87}
  (2013) 055019, [\href{http://xxx.lanl.gov/abs/1212.3307}{{\tt
  arXiv:1212.3307}}].

\bibitem{Lopez-Honorez:2013wla}
L.~Lopez-Honorez and L.~Merlo,  Phys.Lett. {\bf B722} (2013) 135--143,
  [\href{http://xxx.lanl.gov/abs/1303.1087}{{\tt arXiv:1303.1087}}].

\bibitem{Isidori:2010kg}
G.~Isidori, Y.~Nir, and G.~Perez,  Ann. Rev. Nucl. Part. Sci. {\bf 60} (2010)
  355, [\href{http://xxx.lanl.gov/abs/1002.0900}{{\tt arXiv:1002.0900}}].

\bibitem{Anselm:1996jm}
A.~Anselm and Z.~Berezhiani,  Nucl.Phys. {\bf B484} (1997) 97--123,
  [\href{http://xxx.lanl.gov/abs/hep-ph/9605400}{{\tt hep-ph/9605400}}].

\bibitem{Barbieri:1999km}
R.~Barbieri, L.~J. Hall, G.~L. Kane, and G.~G. Ross,
  \href{http://xxx.lanl.gov/abs/hep-ph/9901228}{{\tt hep-ph/9901228}}.

\bibitem{Berezhiani:2001mh}
Z.~Berezhiani and A.~Rossi,  Nucl.Phys.Proc.Suppl. {\bf 101} (2001) 410--420,
  [\href{http://xxx.lanl.gov/abs/hep-ph/0107054}{{\tt hep-ph/0107054}}].

\bibitem{Harrison:2005dj}
P.~Harrison and W.~Scott,  Phys.Lett. {\bf B628} (2005) 93,
  [\href{http://xxx.lanl.gov/abs/hep-ph/0508012}{{\tt hep-ph/0508012}}].

\bibitem{Nardi:2011st}
E.~Nardi,  Phys.Rev. {\bf D84} (2011) 036008,
  [\href{http://xxx.lanl.gov/abs/1105.1770}{{\tt arXiv:1105.1770}}].

\bibitem{Espinosa:2012uu}
J.~R. Espinosa, C.~S. Fong, and E.~Nardi,  JHEP {\bf 1302} (2013) 137,
  [\href{http://xxx.lanl.gov/abs/1211.6428}{{\tt arXiv:1211.6428}}].

\bibitem{Altarelli:2010gt}
G.~Altarelli and F.~Feruglio,  Rev. Mod. Phys. {\bf 82} (2010) 2701--2729,
  [\href{http://xxx.lanl.gov/abs/1002.0211}{{\tt arXiv:1002.0211}}].

\bibitem{Ishimori:2010au}
H.~Ishimori {\em et.~al.},  Prog. Theor. Phys. Suppl. {\bf 183} (2010) 1--163,
  [\href{http://xxx.lanl.gov/abs/1003.3552}{{\tt arXiv:1003.3552}}].

\bibitem{Grimus:2011fk}
W.~Grimus and P.~O. Ludl,  \href{http://xxx.lanl.gov/abs/1110.6376}{{\tt
  arXiv:1110.6376}}.

\bibitem{Altarelli:2012bn}
G.~Altarelli, F.~Feruglio, L.~Merlo, and E.~Stamou,  JHEP {\bf 1208} (2012)
  021, [\href{http://xxx.lanl.gov/abs/1205.4670}{{\tt arXiv:1205.4670}}].

\bibitem{Altarelli:2012ss}
G.~Altarelli, F.~Feruglio, and L.~Merlo,  Fortsch.Phys. {\bf 61} (2013)
  507--534, [\href{http://xxx.lanl.gov/abs/1205.5133}{{\tt arXiv:1205.5133}}].

\bibitem{Bazzocchi:2012st}
F.~Bazzocchi and L.~Merlo,  Fortsch.Phys. {\bf 61} (2013) 571--596,
  [\href{http://xxx.lanl.gov/abs/1205.5135}{{\tt arXiv:1205.5135}}].

\bibitem{Casas:2001sr}
J.~Casas and A.~Ibarra,  Nucl.Phys. {\bf B618} (2001) 171--204,
  [\href{http://xxx.lanl.gov/abs/hep-ph/0103065}{{\tt hep-ph/0103065}}].

\bibitem{Feldmann:2010yp}
T.~Feldmann,  JHEP {\bf 04} (2011) 043,
  [\href{http://xxx.lanl.gov/abs/1010.2116}{{\tt arXiv:1010.2116}}].

\bibitem{Guadagnoli:2011id}
D.~Guadagnoli, R.~N. Mohapatra, and I.~Sung,  JHEP {\bf 04} (2011) 093,
  [\href{http://xxx.lanl.gov/abs/1103.4170}{{\tt arXiv:1103.4170}}].

\bibitem{Buras:2011zb}
A.~J. Buras, L.~Merlo, and E.~Stamou,  JHEP {\bf 1108} (2011) 124,
  [\href{http://xxx.lanl.gov/abs/1105.5146}{{\tt arXiv:1105.5146}}].

\bibitem{Jenkins:2007ip}
E.~E. Jenkins and A.~V. Manohar,  Nucl.Phys. {\bf B792} (2008) 187--205,
  [\href{http://xxx.lanl.gov/abs/0706.4313}{{\tt arXiv:0706.4313}}].

\bibitem{Jenkins:2009dy}
E.~E. Jenkins and A.~V. Manohar,  JHEP {\bf 0910} (2009) 094,
  [\href{http://xxx.lanl.gov/abs/0907.4763}{{\tt arXiv:0907.4763}}].

\bibitem{Broncano:2002rw}
A.~Broncano, M.~Gavela, and E.~E. Jenkins,  Phys.Lett. {\bf B552} (2003)
  177--184, [\href{http://xxx.lanl.gov/abs/hep-ph/0210271}{{\tt
  hep-ph/0210271}}].

\bibitem{Alonso:2013nca}
R.~Alonso, M.~Gavela, G.~Isidori, and L.~Maiani,
  \href{http://xxx.lanl.gov/abs/1306.5927}{{\tt arXiv:1306.5927}}.

\end{thebibliography}
\end{document}